\documentclass[fleqn,twoside]{elsart}
\usepackage{cite,graphicx,amsmath,amsfonts,amssymb,mathrsfs,epic,eepic}
\usepackage[figuresright]{rotating}

\allowdisplaybreaks[1]

\journal{Physics Letters B}

\begin{document}
\begin{frontmatter}

\title{Effective Neutrino Mixing and Oscillations in Dense Matter}

\author[KTH]{Mattias Blennow},
\ead{mbl@theophys.kth.se}
\author[KTH]{Tommy Ohlsson}
\ead{tommy@theophys.kth.se}

\address[KTH]{Division of Mathematical Physics, Department of Physics,
Royal Institute of Technology (KTH) - AlbaNova University Center,
Roslagstullsbacken 11, 106 91 Stockholm, Sweden}

\begin{abstract}
We investigate the effective case of two-flavor neutrino oscillations
in infinitely dense matter by using a perturbative approach. We begin
by briefly summarizing the conditions for the three-flavor neutrino
oscillation probabilities to take on the same form as the
corresponding two-flavor probabilities. Then, we proceed with the
infinitely dense matter calculations. Finally, we study the validity
of the approximation of infinitely dense matter when the effective
matter potential is large, but not infinite, this is done by using
both analytic and numeric methods.
\end{abstract}
\end{frontmatter}

\date{\today}

\maketitle

\section{Introduction}

The large experimental collaborations
\cite{Fukuda:1998mi,Ashie:2004mr,SKNobel,Ahmad:2002jz,Ahmed:2003kj,SNONobel,Ahn:2002up,Eguchi:2002dm,Araki:2004mb,KamLANDNobel}
in neutrino oscillation physics have successfully used effective
two-flavor neutrino oscillation formulas in their analyzes in order to
determine the fundamental neutrino oscillation parameters. However, we
now know that there exists (at least) three active neutrino flavors in
Nature, which, in principle, means that the analyzes have to be
carried out using three-flavor neutrino oscillation formulas. Thus, it
is important to carefully examine the effective two-flavor formulas in
order to determine their validity in different situations, especially
since neutrino oscillation physics has now entered the era of
precision measurements. For example, the Super-Kamiokande
\cite{Fukuda:1998mi,Ashie:2004mr,SKNobel}, SNO
\cite{Ahmad:2002jz,Ahmed:2003kj,SNONobel}, K2K \cite{Ahn:2002up}, and
KamLAND \cite{Eguchi:2002dm,Araki:2004mb,KamLANDNobel} collaborations have significantly pinned down
the errors on the fundamental parameters and future results of the
above mentioned experiments as well as other long-baseline experiments
will continue to decrease the uncertainties of these parameters.

Furthermore, it is known that the presence of matter can result in
large alterations in the behavior of neutrino oscillations
\cite{Wolfenstein:1978ue}. An example of this is the
Mikheyev--Smirnov--Wolfenstein (MSW) effect
\cite{Wolfenstein:1978ue,Mikheev:1985gs}, which is the most plausible
description of the oscillations of solar electron neutrinos into
neutrinos of different flavors. The matter effects are proportional to
the energy of the neutrinos as well as the matter density. Thus, for
large neutrino energy and dense matter, the matter effects become
important. Note that, in Ref.~\cite{Freund:2001pn}, approximate
mappings between the three-flavor neutrino parameters in vacuum and
matter have been obtained. In addition, in
Refs. \cite{Barger:1980tf,Kim:1987vg,Zaglauer:1988gz,Ohlsson:1999xb,Xing:2000gg,Ohlsson:2001vp,Kimura:2002hb,Harrison:2003fi},
exact treatments of three-flavor neutrino oscillations in constant
matter density have been performed, whereas, in
Refs. \cite{Akhmedov:1998xq,Peres:1999yi,Akhmedov:2001kd,Peres:2002ri,Yasuda:1999uv,Freund:1999gy,Mocioiu:2001jy,Arafune:1997hd,Brahmachari:2003bk,Cervera:2000kp,Freund:2001pn,Freund:2001ui,Barger:2001yr,Akhmedov:2004ny},
different analytic approximations of three-flavor neutrino oscillation
probability formulas have been investigated for both constant and
varying matter density.

In this Letter, we study the matter effects on neutrino mixing and
oscillations in the limit where the matter density becomes infinite
and the resulting neutrino oscillation probabilities are actual
two-flavor formulas. We also investigate how well these two-flavor
formulas reproduce the actual neutrino oscillation probabilities when
neutrinos propagate through matter with large, but not infinite,
matter density.

\section{Effective Two-Flavor Formulas}  

With three neutrino flavors, there are six fundamental parameters
which influence neutrino oscillations, two mass squared differences
($\Delta m_{21}^2$ and $\Delta m_{31}^2$), three mixing angles
($\theta_{12}$, $\theta_{23}$, and $\theta_{13}$), and one {\it
CP}-violating phase ($\delta$). For the leptonic mixing matrix $U$, we
adopt the standard parameterization given in
Ref.~\cite{Eidelman:2004wy}. In addition, we introduce the mass
hierarchy parameter $\alpha \equiv \Delta m_{21}^2/\Delta m_{31}^2$ as
well as the parameter $\Delta \equiv \Delta m_{31}^2/(2E)$, where $E$
is the neutrino energy. We also use the abbreviations $s_{ij} \equiv
\sin \theta_{ij}$, $c_{ij} \equiv \cos \theta_{ij}$, $s_\delta \equiv
\sin\delta$, and $c_\delta \equiv \cos\delta$.

In some special cases, it is possible to write one or more of the
neutrino oscillation probabilities as effective two-flavor formulas,
\emph{i.e.}, the probabilities can be written as
\begin{equation}
\label{eq:2flav}
P_{\alpha\beta} = \delta_{\alpha\beta} + (1-2\delta_{\alpha\beta}) 
\sin^2(2\theta)\sin^2\left(\frac{\Delta m^2}{4E}L\right),
\end{equation}
where $P_{\alpha\beta}$ is the probability of the transition
$\nu_\alpha \rightarrow \nu_\beta$, $\sin^2(2\theta)$ defines the
oscillation amplitude, $\Delta m^2$ defines the oscillation frequency,
and $L$ is the length of the path travelled by the neutrinos. In the
case of three-flavor neutrino oscillations in vacuum, the survival
probability $P_{\alpha\alpha}$ is given by
\begin{equation}
P_{\alpha\alpha} = 1 - 4 \sum_{1\leq i < j \leq 3} 
|U_{\alpha i}|^2 |U_{\alpha j}|^2
\sin^2\left(\frac{\Delta m_{ij}^2}{4E}L\right).
\end{equation}
This probability takes the form of Eq.~(\ref{eq:2flav}) if any of the
elements $U_{\alpha i}$ of the leptonic mixing matrix or any of the
mass squared differences equals zero. The same argument holds for the
neutrino oscillation probability $P_{\alpha\beta}$. If one of the
conditions is satisfied, then the leptonic mixing matrix can be made
real and there is no {\it CP}-violation in neutrino oscillations
(Majorana phases, which do not influence neutrino oscillations, could
still be present if neutrinos are Majorana particles). In that case,
the neutrino oscillation probability is given by
\begin{equation}
P_{\alpha\beta} = - 4 \sum_{1\leq i<j \leq 3} 
U_{\alpha i}^*U_{\beta i}U_{\alpha j}
U_{\beta j}^* \sin^2\left(\frac{\Delta m_{ij}^2}{4E}L\right).
\end{equation}
Note that if $U_{\alpha i} = 0$, then only the oscillations involving the
neutrino flavor $\nu_\alpha$ will become effective two-flavor cases.

The two-flavor cases mentioned above are realized if we make either of
the approximations $\theta_{13} \rightarrow 0$ or $\alpha \rightarrow
0$, see Table \ref{tab:2flav} for amplitudes and frequencies. These
approximations can be used when analyzing neutrino oscillation
experiments neglecting matter effects, for example, the oscillations
of electron anti-neutrinos of reactor experiments such as KamLAND
\cite{Eguchi:2002dm,Araki:2004mb,KamLANDNobel} ($\theta_{13}
\rightarrow 0$) and oscillations of atmospheric neutrinos in
experiments such as Super-Kamiokande
\cite{Fukuda:1998mi,Ashie:2004mr,SKNobel} ($\alpha \rightarrow 0$).
\begin{table}
\begin{center}
\begin{tabular}{|l|l|l|}
\hline
$\boldsymbol{P_{\alpha\beta}}$ & $\boldsymbol{\sin^2(2\theta)}$ &
$\boldsymbol{\Delta m^2}$ \\
\hline
\multicolumn{3}{|l|}{$\boldsymbol{\theta_{13}=0}$} \\
\hline
$P_{ee}$ & $\sin^2(2\theta_{12})$ & $\Delta m_{21}^2$ \\
$P_{e\mu} = P_{\mu e}$ & $c_{23}^2 \sin^2(2\theta_{12})$ & $\Delta m_{21}^2$ \\
$P_{e\tau} = P_{\tau e}$ & $s_{23}^2 \sin^2(2\theta_{12})$ & 
$\Delta m_{21}^2$ \\
\hline
\multicolumn{3}{|l|}{$\boldsymbol{\alpha=0}$} \\
\hline
$P_{e\mu}$ & $s_{23}^2\sin^2(2\theta_{13})$ & 
$\Delta m_{31}^2 = \Delta m_{32}^2$ \\
$P_{\mu\tau}$ & $c_{13}^4 \sin^2(2\theta_{23})$ & 
$\Delta m_{31}^2 = \Delta m_{32}^2$ \\
\hline
\end{tabular}
\caption{The neutrino oscillation amplitudes and frequencies for
different two-flavor formulas in vacuum. In the $\alpha = 0$ case, all
channels are two-flavor channels and the amplitudes and frequencies
can be deduced from the channels in this table (see
Ref.~\cite{Akhmedov:2004ny}).}
\label{tab:2flav}
\end{center}
\end{table}

The above argument also holds for neutrinos propagating through matter
of constant density if exchanging the fundamental neutrino parameters
for their effective counterparts in matter. This gives us another
important two-flavor case, namely the limit of an effective matter
potential approaching infinity, where all neutrino oscillation
probabilities reduce to the two-flavor form of
Eq.~(\ref{eq:2flav}). The remainder of this Letter is devoted to study
this limit in detail.

\section{Mixing in Dense Matter}

In matter, the Hamiltonian of three flavor neutrino evolution consists
of two parts, the kinematic term $H_k = \Delta U {\rm diag}(0,\alpha,
1) U^\dagger$ and the interaction term $H_m = {\rm diag}(V, 0, 0)$,
where $V = \sqrt 2 G_F n_e$, $G_F$ is the Fermi coupling constant, and
$n_e$ is the electron number density, resulting from coherent
forward-scattering of neutrinos. When $2EV \gg \Delta m_{31}^2$, the
term $H_k$ can be thought of as a perturbation to the term
$H_m$. Since $H_m$ has two degenerate eigenvectors, we can use
degenerate perturbation theory to find approximate eigenvectors and
eigenvalues of the total Hamiltonian. The degenerate sector of $H_m$
is spanned by $\nu_\mu = (0,1,0) ^T$ and $\nu_\tau = (0, 0, 1)^T$. In
this sector, the Hamiltonian is of the form
\begin{equation}
H_d = \left(\begin{array}{cc}a & b^* \\ b & d\end{array}\right),
\label{eq:Hd}
\end{equation}
where $a$, $b$, and $d$ depend on the fundamental neutrino parameters
as well as the effective matter potential $V$ and the neutrino energy
$E$.

\begin{figure}
\begin{center}
\includegraphics[width=7cm]{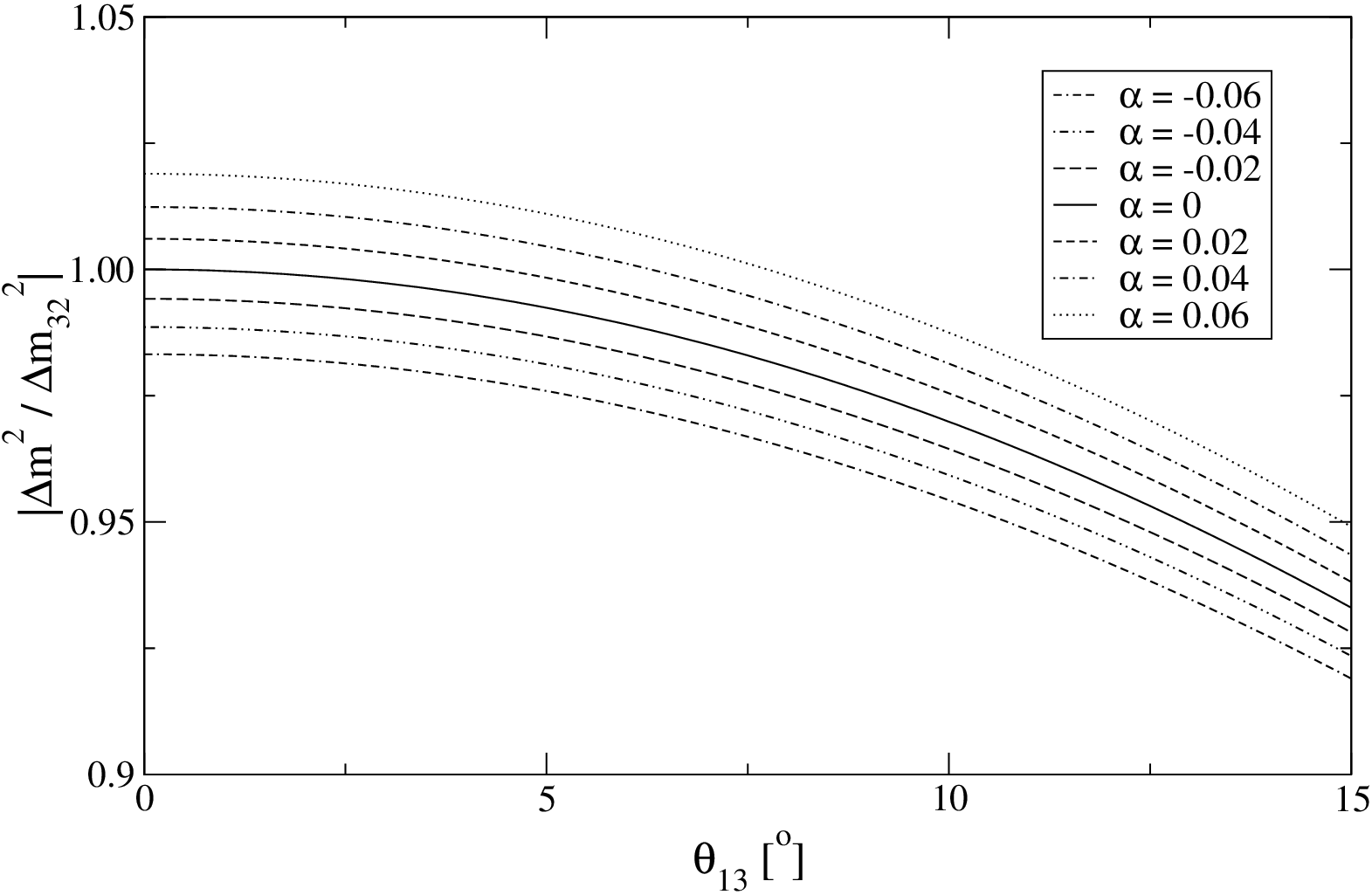}
\end{center}
\caption{The quotient $|\Delta m_\infty^2 / \Delta m_{32}^2|$ as a
function of the leptonic mixing angle $\theta_{13}$ for different
values of the mass hierarchy parameter $\alpha$. In this plot, the
mixing angle $\theta_{12}$ has been set to $33^\circ$.}
\label{fig:m2}
\end{figure}

In the case of two-flavor neutrino oscillations in vacuum, the
Hamiltonian takes the form
\begin{equation}
H = \frac{\Delta m^2}{4E} \left(\begin{array}{cc}
-\cos(2\theta) & \sin(2\theta) \\ \sin(2\theta) & \cos(2\theta)
\end{array}\right), \label{eq:H2f}
\end{equation}
where $\Delta m^2$ and $\theta$ are the mass squared difference and
mixing angle in the two-flavor scenario, respectively. Defining
$\theta_\infty$ as the effective mixing angle in the degenerate sector
of $H_m$ and $\Delta m_\infty^2$ as the effective mass squared
difference, by comparing Eqs.~(\ref{eq:Hd}) and (\ref{eq:H2f}), we
obtain
\begin{eqnarray}
\sin^2(2\theta_\infty) &=& \frac{4|b|^2}{(a-d)^2+4|b|^2} \nonumber \\
&=& 
4\frac{\{s_{23}c_{23}[c_{13}^2-\alpha(c_{12}^2-s_{12}^2s_{13}^2)]-
\alpha s_{12}c_{12}s_{13}c_\delta \cos(2\theta_{23})\}^2}
{[c_{13}^2-\alpha(c_{12}^2-s_{13}^2s_{12}^2)]^2
+4\alpha c_{12}^2s_{12}^2s_{13}^2} \nonumber \\
& &+ 4\frac{\alpha^2 s_\delta^2 s_{12}^2c_{12}^2s_{13}^2}
{[c_{13}^2-\alpha(c_{12}^2-s_{13}^2s_{12}^2)]^2
+4\alpha c_{12}^2s_{12}^2s_{13}^2},\label{eq:effamp} \\
\Delta m_\infty^2 & = & 2E \sqrt{(a-d)^2 + 4|b|^2} \nonumber \\
&=&
 |\Delta m_{31}^2| \sqrt{c_{13}^4 -2\alpha c_{13}^2(c_{12}^2-s_{13}^2 s_{12}^2)
+ \alpha^2(1-c_{13}^2 s_{12}^2)^2} \nonumber \\
&=&
|\Delta m_{31}^2| \sqrt{[c_{13}^2-\alpha(c_{12}^2-s_{13}^2s_{12}^2)]^2
+4\alpha c_{12}^2s_{12}^2s_{13}^2}. \label{eq:effdM}
\end{eqnarray}
We note that even for $\theta_{13} = 0$ or $\alpha = 0$, the effective
two flavor parameters $\Delta m_\infty^2$ and $\theta_\infty$ are not
equal to $\Delta m_{32}^2$ and $\theta_{23}$, which might have been
expected. However, for reasonable values of the fundamental neutrino
parameters, $\theta_\infty$ does not deviate significantly from
$\theta_{23}$. In Fig.~\ref{fig:m2}, the quotient $|\Delta m_\infty^2 /
\Delta m_{32}^2|$ is plotted as a function of $\theta_{13}$ for
different $\alpha$.
From this figure, we conclude that the sign of the correction to the
approximation $\Delta m_\infty^2 = |\Delta m_{32}^2|$ depends on the
values of the parameters $\theta_{13}$ and $\alpha$. It should be
noted that if the absolute value of the fundamental parameter $\Delta
m_{32}^2$ and the effective parameter $\Delta m_\infty^2$ were to be
measured (\emph{i.e.}, $\Delta m_\infty^2$ determined in an experiment
where the matter is extremely dense and $|\Delta m_{32}^2|$ is
determined by vacuum or close to vacuum experiments), then the
quotient $|\Delta m_\infty^2/\Delta m_{32}^2|$ could provide valuable
information on the mass hierarchy (\emph{i.e.}, the sign of $\alpha$)
and the leptonic mixing angle $\theta_{13}$. It is also interesting to
observe that for $\alpha \geq 0$ (normal mass hierarchy), there exists
a value of $\theta_{13}$ such that $\Delta m_\infty^2 = |\Delta
m_{32}^2|$.

\begin{figure}
\begin{center}
\includegraphics[width=6.5cm]{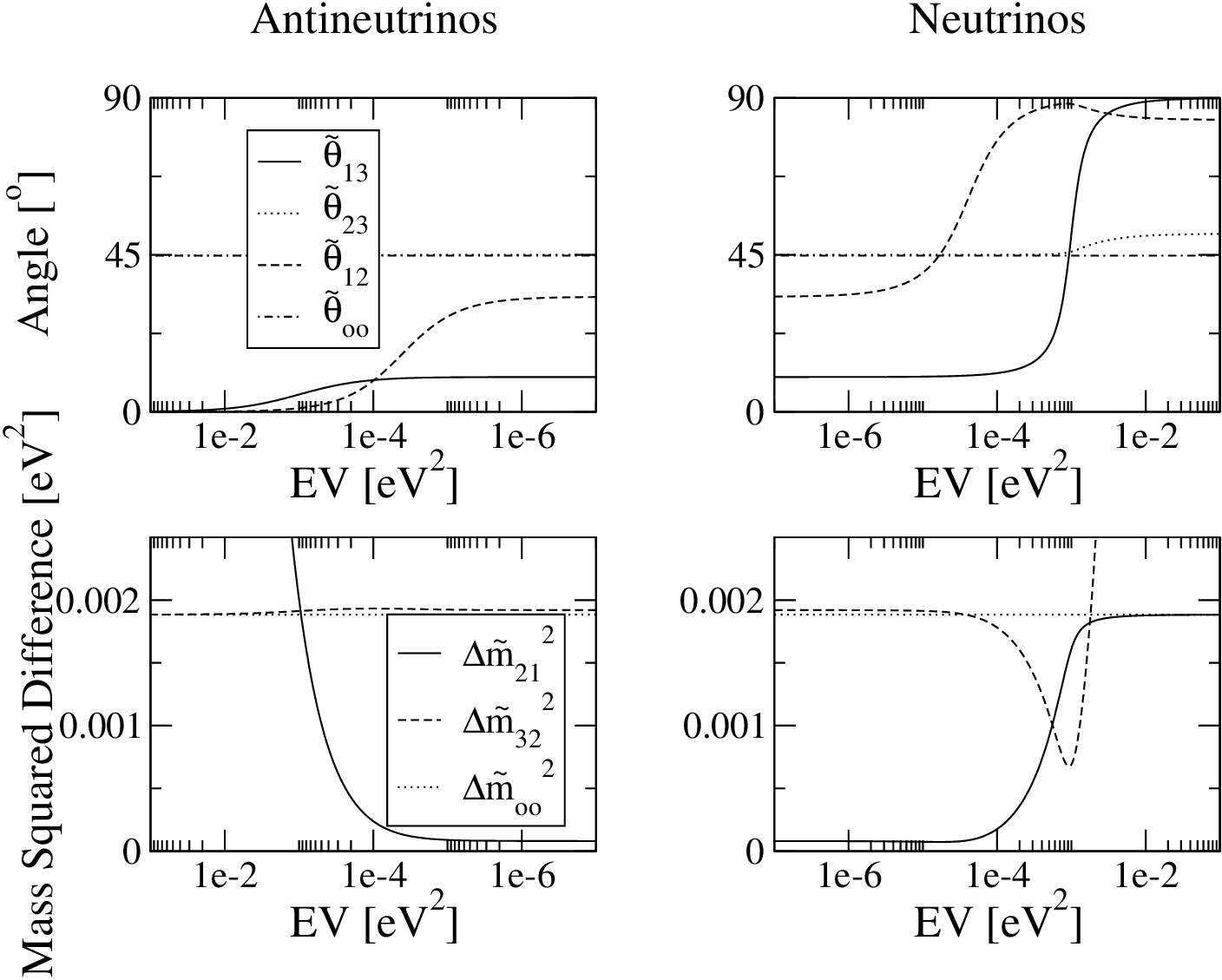}
\end{center}
\caption{Numerical results for the effective neutrino parameters in
matter as a function of $EV$ for a normal mass hierarchy with the
vacuum parameters $\Delta m_{21}^2 = 8\cdot 10^{-5}$ eV$^2$, $\Delta
m_{31}^2 = 2\cdot 10^{-3}$ eV$^2$, $\theta_{12} = 33^\circ$,
$\theta_{23} = 45^\circ$, $\theta_{13} = 10^\circ$, and $\delta = 0$.}
\label{fig:effparam}
\end{figure}

In Fig.~\ref{fig:effparam}, we plot the effective neutrino parameters in
matter as a function of the product $EV$.
For neutrinos at large values of $EV$, the effective mixing angles
$\tilde \theta_{12}$ and $\tilde \theta_{23}$, where the tilde denotes
that these are the effective matter parameters, do not play a very
important role, since $\tilde \theta_{13} \rightarrow 90^\circ$ when
$EV \rightarrow \infty$. Instead, in this case, the important mixing
angle is the mixing between the first and second eigenstates, which
for $\tilde \theta_{13} = 90^\circ$ becomes dependent on the the
effective mixing angles $\tilde \theta_{12}$ and $\tilde \theta_{23}$,
as well as the effective phase $\tilde \delta$, this angle is clearly
the angle $\theta_\infty$.

As can be seen from Fig.~\ref{fig:effparam}, we may consider
essentially three different regions for the parameter $EV$. The first
region is $2EV \ll \Delta m_{21}^2$, where the fundamental neutrino
parameters are essentially equal to the vacuum parameters, the second
region is $2EV \gg \Delta m_{31}^2$, where $\nu_e$ is essentially an
eigenstate to the Hamiltonian and there is two-flavor neutrino mixing
between $\nu_\mu$ and $\nu_\tau$. The last region is the region
inbetween the other two, where resonance phenomena occur. In general,
the results for an inverted mass hierarchy are similar to those
of a normal mass hierarchy.

Observe that for atmospheric neutrinos passing through the Earth with
a characteristic energy of $E \simeq 1$ GeV, the product $EV$ is
typically in the upper part of the resonance region in which none of
the effective mass squared differences, nor any of the effective
mixing angles, are small. This effect, which is studied in, for
example, Ref.~\cite{Freund:1999vc} (for a recent study, see
Ref.~\cite{Gandhi:2004md}), should be taken into account when
analyzing data from atmospheric neutrino experiments.

\section{Accuracy of the Two-Flavor Approximation}

The accuracy of the two-flavor approximation in the limit $EV
\rightarrow \infty$ is dependent on how close the leptonic matter
mixing angle $\tilde \theta_{13}$ is to $90^\circ$, or in other words,
the value of $\tilde c_{13}$. The correction to $\nu_3 = \nu_e$ is
given by
\begin{equation}
\nu_3-\nu_e \simeq \frac{H_{e\mu}\nu_\mu + H_{e\tau} \nu_\tau}{V}
\end{equation}
in first order non-degenerate perturbation theory (which can be used,
since $\nu_e$ is a non-degenerate eigenstate of $H_m$), where
$H_{e\alpha}$ is the $e$-$\alpha$ element of $H_k$. From this
calculation follows that to first order in $\Delta m_{31}^2/(EV)$, the
quantity $\tilde c_{13}$ is given by
\begin{equation}
\label{eq:c13eff}
\tilde c_{13} \simeq \frac{\Delta m_{31}^2}{2EV} \sqrt{\frac 14 \sin^2(2\theta_{13})
-2\alpha s_{13}^2c_{13}^2 s_{12}^2 + 
\alpha^2 c_{13}^2 s_{12}^2 (c_{12}^2 + c_{13}^2 s_{12}^2)}.
\end{equation}

\begin{figure}
\begin{center}
\includegraphics[width=7cm]{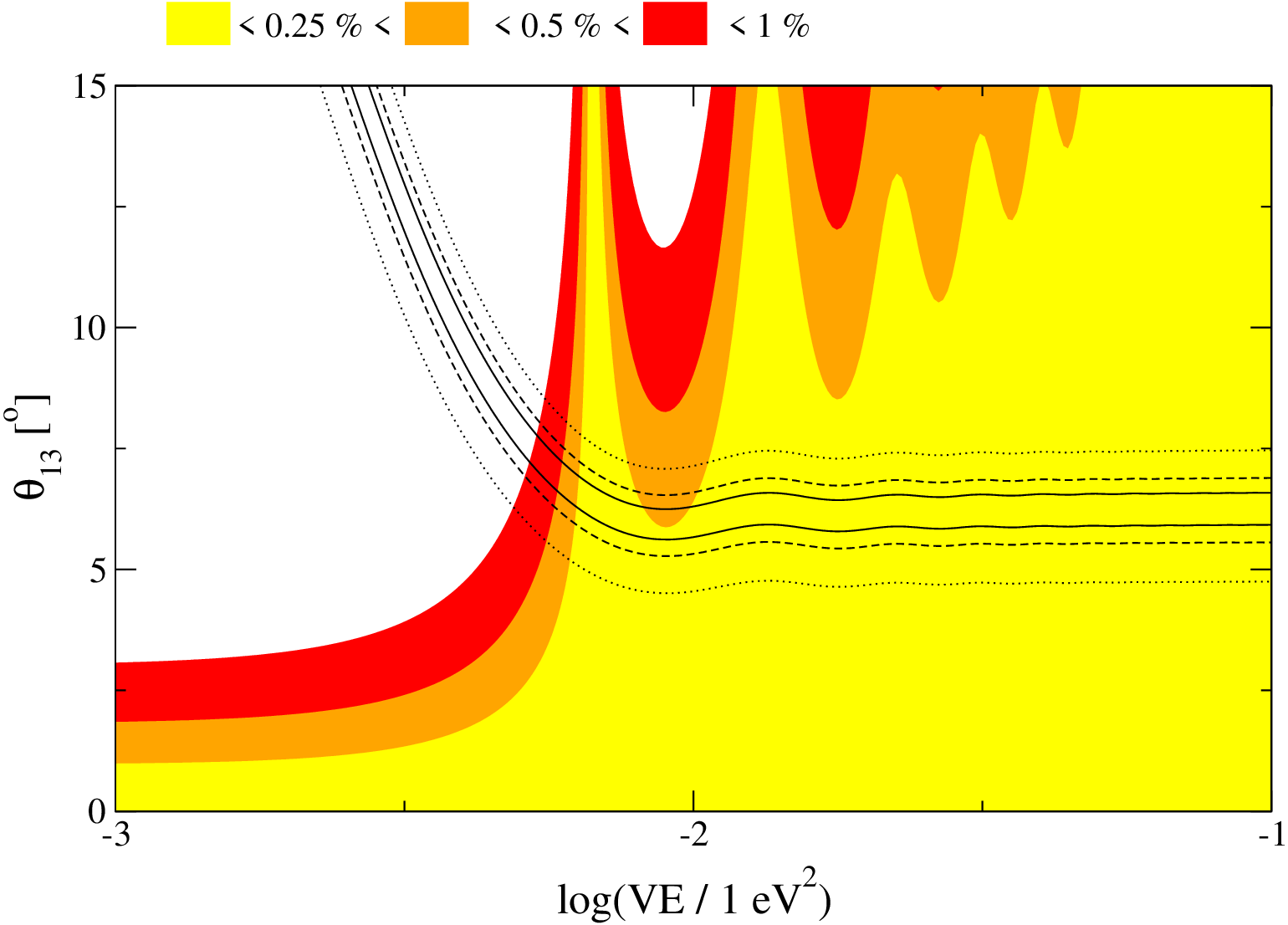} \\[.5cm]
\includegraphics[width=7cm]{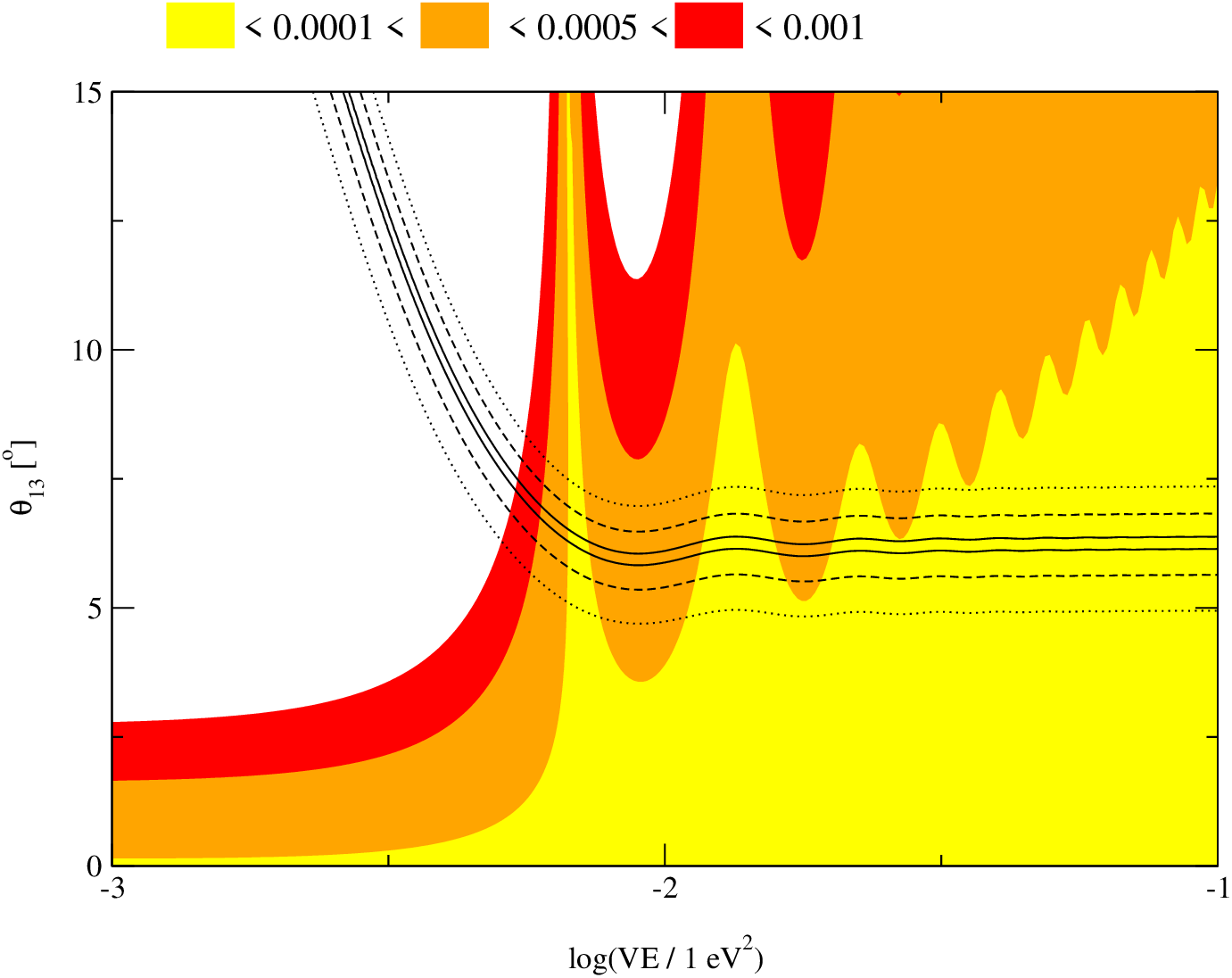}
\end{center}
\caption{The relative (upper panel) and absolute (lower panel)
accuracy of using the approximation $EV \rightarrow \infty$ in the
neutrino oscillation channel $1-P_{\mu\mu}$ for $L/E = 7000$ km / 50
GeV. The fundamental neutrino parameters have been set to $\Delta
m_{21}^2 = 8\cdot 10^{-5}$ eV$^2$, $\Delta m_{31}^2 = 2\cdot
10^{-3}$~eV$^2$, $\theta_{12} = 33^\circ$ $\theta_{23} = 45^\circ$,
and $\delta = 0$. The colored regions correspond to different values
of the relative accuracy $A_{\rm rel} = |P_{\mu\mu,\rm
num}-P_{\mu\mu,\rm app}|/(1-P_{\mu\mu,\rm num})$ and absolute accuracy
$A_{\rm abs} = |P_{\mu\mu,\rm num}-P_{\mu\mu,\rm app}|$ as indicated
by the legends above the panels. The curves corresponds to the
isocontours of the accuracy of the two flavor approximation $\theta =
\theta_{23}$ and $\Delta m^2 = \Delta m_{32}^2$. The solid curves are
the $A_{\rm rel} = 0.25$~\% isocontours, the dashed curves are the
$A_{\rm rel} = 0.5$~\% isocontours, and the dotted curves are the
$A_{\rm rel} = 1$~\% isocontours in the relative accuracy plot. In the
absolute accuracy plot, the solid curves are the $A_{\rm abs} =
0.0001$ isocontours, the dashed curves are the $A_{\rm abs} = 0.0005$
isocontours, and the dotted curves are the $A_{\rm abs} = 0.001$
isocontours.}
\label{fig:rel7}
\end{figure}

\begin{figure}
\begin{center}
\includegraphics[width=7cm]{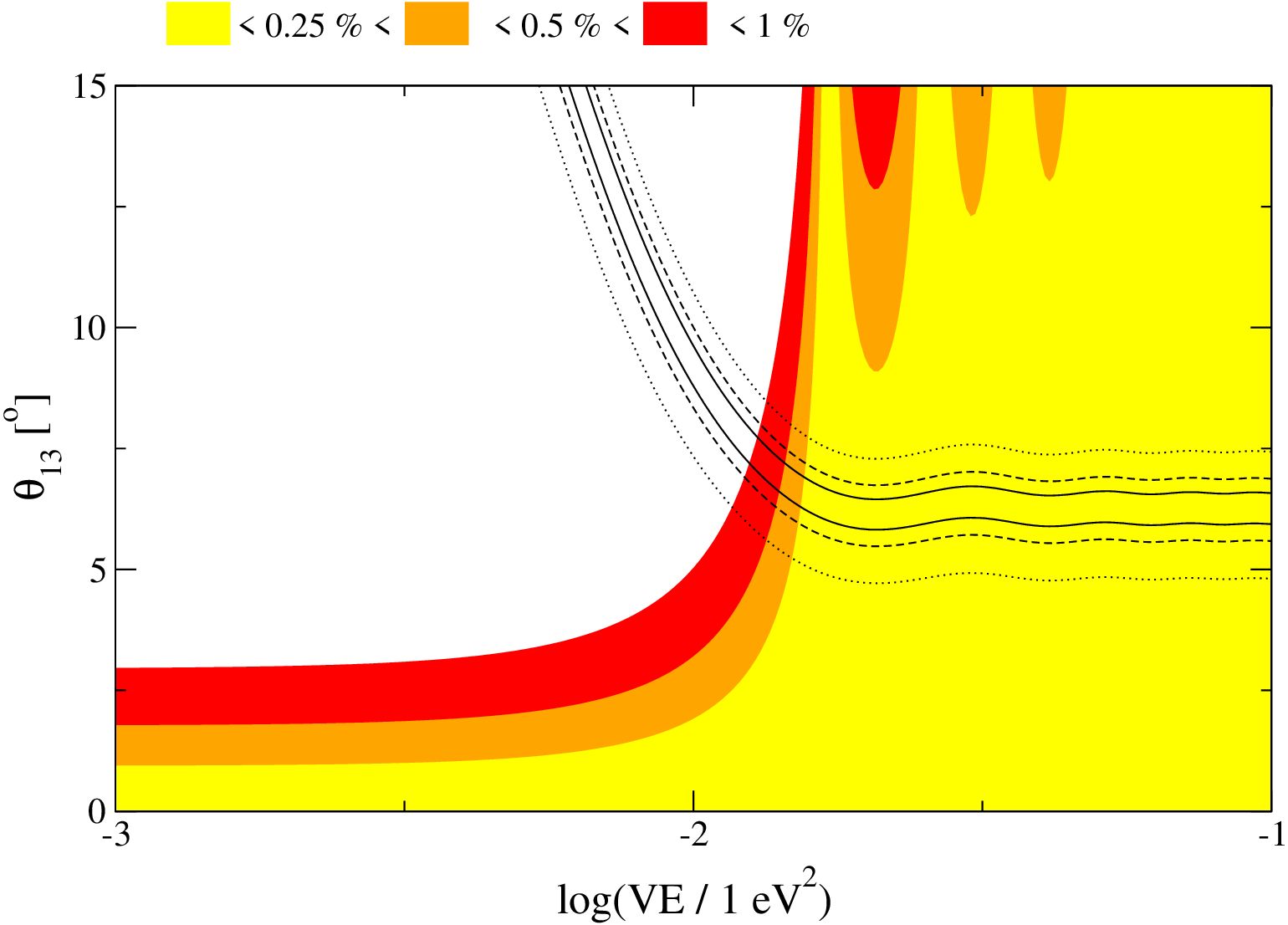} \\[.5cm]
\includegraphics[width=7cm]{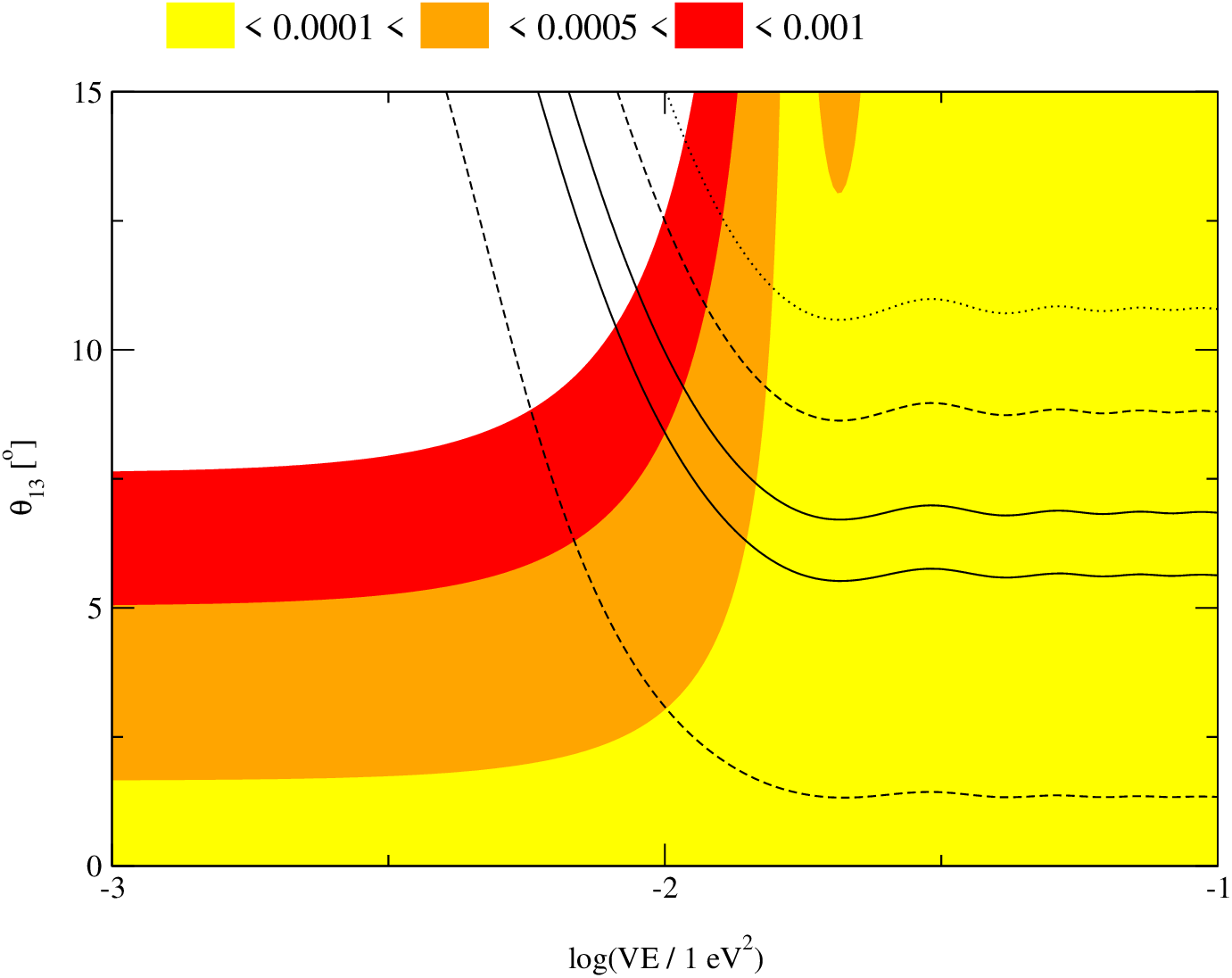}
\end{center}
\caption{The relative (upper panel) and absolute (lower panel)
accuracy of using the approximation $EV \rightarrow \infty$ in the
neutrino oscillation channel $1-P_{\mu\mu}$ for $L/E = 3000$ km / 50
GeV. The areas and curves are the same as described in the figure
caption of Fig.~\ref{fig:rel7}.}
\label{fig:rel3}
\end{figure}

In Figs.~\ref{fig:rel7} and \ref{fig:rel3}, we plot the relative and
absolute accuracies of the large matter density approximation for the
probability $P$ of a $\nu_\mu$ to oscillate into some other flavor in
the limit $EV \rightarrow \infty$ for two different values of $L/E$ as
a function of $EV$ and $\theta_{13}$. Since this oscillation
probability is given by $P = 1-P_{\mu\mu}$, the relative accuracy is
given by $A_{\rm rel} = A_{\rm abs} / P = A_{\rm abs}/(1-P_{\mu\mu})$
where $A_{\rm abs}$ is the absolute accuracy. As for all relative
accuracies, this definition clearly has a problem for $P = 0$. Since
$P_{\mu\mu}$ is close to one in the regime studied in the plots, the
relative accuracy for $P_{\mu\mu}$ is approximately the same as the
absolute accuracy.
The values used for $L/E$ correspond roughly to values that might be
used for future long-baseline experiments and neutrino
factories. Note that these values are smaller than the oscillation length
determined by $\Delta m_{31}^2$.

The reason why the approximation becomes worse at large values of
$\theta_{13}$ is the fact that if $\theta_{13}$ is large, then $\tilde
\theta_{13}$ will approach $90^\circ$ slowly when $EV \rightarrow
\infty$ compared to when $\theta_{13}$ is small, see
Eq.~(\ref{eq:c13eff}). Thus, larger values of $EV$ will be required
before the approximation $EV \rightarrow \infty$ becomes valid. For
$EV \rightarrow \infty$, the region in which the two-flavor neutrino
approximation $\Delta m^2 = \Delta m_{32}^2$ and $\theta =
\theta_{23}$ is accurate is the region in which $\theta_{13}$ obtains
the value for which $\Delta m_\infty \simeq |\Delta m_{32}^2|$. This is
to be expected, since $\theta_\infty \simeq \theta_{23}$. When
studying Figs.~\ref{fig:rel7} and \ref{fig:rel3}, one should keep in
mind that the $\nu_\mu$ survival probability $P_{\mu\mu}$ is larger
for $L/E = 7000$ km / 50 GeV than for $L/E = 3000$ km / 50 GeV and
that we have plotted the relative accuracy of the approximation.

The reason why the two-flavor neutrino approximation $EV \rightarrow
\infty$ seems to be a good approximation even for small values of $EV$
when $\theta_{13}$ is small has to do with the fact that for small
values of $L/E$ and $\theta_{13} = 0$, this approximation gives an
error which is of the order $\alpha^2$ to the $(L/E)^2$ dependence of
the probability $P_{\mu\mu}$ when making a series expansion in the
quantity $L/E$.

Presently, the number of experiments in which the approximation of $EV
\rightarrow \infty$ can be justified is somewhat limited. One such
example could be a neutrino factory setup with a ``magic'' baseline
($L \simeq 7250$ km) \cite{MagicBaseline} with a neutrino energy of
about 50 GeV. For this energy, the value of $\log(EV/{\rm eV}^2)$
typically varies between $-2.5$ and $-1.8$ along the baseline, which
is basically the middle region of Figs.~\ref{fig:rel7} and
\ref{fig:rel3}. Even though the real Earth matter density profile is
not constant, the $EV \rightarrow \infty$ approximation should be at
least as good as for the case of a constant matter density
corresponding to the lowest density of the Earth matter density
profile. The approximation has no direct application to solar or
supernova neutrinos for which there is a larger $V$ than that given by
the Earth. This is due to the fact that these situations are
well-described by adiabatic evolution and that the neutrino energies
are lower than in the case of a neutrino factory. The approximation
might also be applicable to determine the second order oscillation
effects in high-energy neutrino absorption tomography of the Earth.

\section{Summary and Conclusions}

We have studied effective two-flavor neutrino mixing and oscillations
in the limit $EV \rightarrow \infty$. In this limit, the neutrino
oscillations take the form of two-flavor $\nu_\mu \leftrightarrow
\nu_\tau$ oscillations with the frequency and amplitude given by
Eqs.~(\ref{eq:effamp}) and (\ref{eq:effdM}), respectively. It has been
noted that, if measured, the quotient $|\Delta m_\infty^2/\Delta
m_{32}^2|$ could provide valuable information on the mass hierarchy in
the neutrino sector and the leptonic mixing angle $\theta_{13}$. We
have also examined the validity of using the mentioned two-flavor
formulas for large, but not infinite, matter density. The deviation
from the two-flavor scenario is determined by the deviation of $\tilde
c_{13}$ from zero, which is given by Eq.~(\ref{eq:c13eff}) to first
order in perturbation theory. The validity of the approximation has
also been studied numerically with the results presented in
Figs.~\ref{fig:rel7} and \ref{fig:rel3}.

\section*{Acknowledgments} 
This work was supported by the Swedish Research Council
(Vetenskapsr{\aa}det), Contract Nos.~621-2001-1611, 621-2002-3577, the
G{\"o}ran Gustafsson Foundation, and the Magnus Bergvall Foundation.


\end{document}